\documentclass{article}
\usepackage[utf8]{inputenc}
\usepackage{amsmath}
\usepackage{amsthm}
\usepackage{amssymb}
\usepackage{algpseudocode}
\usepackage{algorithm}
\usepackage[table,xcdraw]{xcolor}
\usepackage{layout}
\usepackage{graphicx}
\usepackage{pgfplots}
\usepackage{placeins}
\usepackage{pgfplots}
\usepackage{tikz}

\pgfplotsset{width=10cm, height=8cm, compat=1.9}

\usepgfplotslibrary{external}
\theoremstyle{definition}

\begin{document}

\thispagestyle{plain}
\hspace{0pt}
\vfill
\begin{center}

    \Large
    \textbf{Identification of Direct Socio-Geographical Price Discrimination}
        
    \vspace{0.4cm}
    \large
    An Empirical Study on iPhones
        
    \vspace{0.4cm}
    \textbf{Davidson Cheng} \\ 
    \text{Department of Mathematics and Computer Science} \\ 
    \text{Colorado College} \\ 
    \text{Colorado Springs, CO, USA} \\ 
    \text{d\_cheng@coloradocollege.edu} \\ 
       
    \vspace{0.9cm}

    \vspace{0.9cm}

\end{center}
\begin{center}
    \textbf{Abstract}
\end{center}
Price discrimination is a practice where firms utilize varying sensitivities to prices among consumers to increase profits. The welfare effects of price discrimination is not agreed on among economists, but identification of such actions may contribute to our standing of firms' pricing behaviors. In this letter, I use econometric tools to analyze whether Apple Inc, one of the largest company in the globe, is practicing price discrimination on the basis of socio-economical and geographical factors. My results indicates that iPhones are significantly (p $<$ 0.01) more expensive in markets where competitions are weak or where Apple has a strong market presence. Furthermore, iPhone prices are likely to increase (p $<$ 0.01) in developing countries/regions or markets with high income inequality.

\vfill
\hspace{0pt}
\pagenumbering{gobble}
\clearpage
\setcounter{page}{1}
\pagenumbering{arabic}
\pagebreak

\section{Introduction}

Naturally, markets contain consumers with different sensitivities towards prices. Firms may choose to take advantage of such varying sensitivity to maximize profit; this practice is called price discrimination or personalized pricing. Apple Inc. is considered by many to have changed the personalized computer and phone industry, it is also once the world’s largest company with market capitalization in the trillions. The iPhone is a popular product from Apple sold across the globe. In fact, the prices of iPhones vary in an interesting fashion across different markets. For example, as of October 2021, the iPhone 13 with 128gb of storage is sold at \$799 in the U.S. and roughly \$1441 in Brazil, an economy with much less mean income. Selling iPhones in Brazil will require additional costs that may otherwise be avoided in the U.S. But when those additional costs are accounted for, are iPhones still priced higher in countries such as Brazil than in the U.S? 

In this letter, I examine Apple Inc’s pricing strategy across countries with various economic status to identify whether apple Inc. is conducting price discrimination in different geographical regions to increase its profit. I first define price discrimination and survey both theoretical and empirical work in the literature. Subsequently, the methodology is outlined and data is introduced. Finally, I present and discuss my results with respect to theories on price discrimination. 

\section{Definition and Background}

 In general, firms conduct price discrimination by offering different prices to customers with different demands. Lower prices for price-sensitive customers and higher prices for those relatively price-insensitive would increase the firm’s overall profit. Pigou has classified price discrimination into three degrees. In the first degree of price discrimination, or pure price discrimination, the firm charges each customer the maximum price they are willing to pay; the second degree is when prices are varied based on quantity of goods purchased; the third degree refers to when different prices are offered to different groups of customers [Pigou, 1920]. McAfee points out that Pigou’s classificatons are flawed [McAFee, 2008], as the second degree do not represent an intermediate step between the first and third degree. In addition, the first and third degrees of price discrimination offer the consumer a singular price while in the second degree of price discrimination the consumer can choose the price by adjusting quantity purchased. A more modern taxonomy of price discrimination includes direct and indirect price discrimination, whereas direct price discrimination offers consumers different prices based on observable features such as geographical location, age, gender, and indirect price discrimination utilizes customers’ unobservable features such as preferences and demands. Despite disagreements in classifications, scholars have agreed that market power and arbitrage difficulties are requisites to conducting price discrimination.
 
In the literature, criticisms to studies that claim to have identified price discrimination speak to the failure to account for differentiated cost. Lott, Russel, and Roberts have pointed out that differences in cost or quality can explain price difference for alleged of price discrimination in gasoline , dinner, and airline prices [Lott, 1991]. Lott, Russel, and Roberts further attempted to provide an explanation for this phenomenon: it is easier to attribute price differences to a firm wielding its monopolistic power than gathering empirical evidences that might account for the additional cost. Two formal definitions of price discrimination that considers differentiated costs have been popularly adopted. The first one, offered by Phlips in 1983, proposes that differences in profits constitutes price discrimination, as in price discrimination exists when the same commodity is sold at $price_1$ and $price_2$, and 

\[price_2 - cost_2 \neq price_1 - cost_1. \tag{1}\]

The second definition follows Stigler’s concept which claims that price discrimination is present when profit markups are unequal. Formally, price discrimination is present whenever 

\[ln(price_2) - ln(price_2) \neq ln(cost_2) - ln(cost_1). \tag{2}\]

The two models are based on different assumptions of the firm’s profit strategy. In the case of non discriminatory pricing, Philip’s measure assumes constant profit while Stigler’s measure assumes constant markup. Clerides have shown that rejection of price discrimination using one of the measure always lead to acceptance in another. Therefore it is critical to pick and justify the the measure before performing any empirical analysis.

\section{Methodology}

Based on the following assumptions, we choose Stigler’s measure over Phlips’ in this case study.

\textbf{Assumption 1:} If iPhone’s pricing is non-discriminatory, it would maintain the same profit markup across variations of iPhones. 

\textbf{Assumption 2:} Prices of iPhones do not dependent on quantity. 

\textbf{Assumption 3:} Costs of iPhones can be linearly estimated using model and market-specific parameters.

Apple have always offered iPhones of different graeds (“Pro”, “S”, “Mini”, “SE”, etc.). If sales of premium iPhone models results in the same profit as the other models, it must be true that more-premium models have a lower profit margin. Regardless of societal and marketing benefits of premium models, this would reduce Apple’s incentive to develop premium iPhone models. However, Apple has always offered iPhones with different storage capacity options with the high-storage models at higher prices. Since 2011, Apple has also been offering iPhones with the affixes “S”, “Plus”, “Pro”, and is continuing to develop premium models. We therefore can assume that the premium iPhone models do not have lower profit markup than the rest, which justifies the first assumption to a certain degree. The second assumption implies non-discriminatorily priced iPhones will maintain equal profit markup when sold at different quantities. This assumption is rather strong as it is well-known that Apple offers one clear-cut price for customers regardless of their demand: there is never “buy two get one 50\% off” in the Apple Store. Companies that offer lower prices to customers with high demand may be subject to indirect or the second degree of price discrimination, as the customers with higher demands are typically more price-sensitive. This practice is also called nonlinear pricing. While Apple do not seem to price iPhones nonlinearly, Apple does allow customers to trade-in their old iPhones to get discounts.  This can be considered a sign for indirect price discrimination, as customers who are willing to trade-in their old phones are likely to be more price-sensitive. The trade-in program is not considered in this study because it is not available in every market. 

Stigler's definition of price discrimination takes in the products' costs as a variable. However, such information is not publicly available, so we make Assumption 3 and apply a linear model  to estimate the cost of iPhones:

\[\hat{cost} = \beta_0 + \beta_{mdl} \times X'_{mdl} + \beta_{mkt} \times X'_{mkt}. \tag{3}\]

The model-specific variables include variables that are independent across markets (storage capacity, features, generation of the phone, etc.). The market-specific variables account for differentiated cost for selling at a different market (taxes, tariffs, operational cost, etc.); these variables are independent of model-specific parameters. Substituting $\hat{cost}$ for $c$ in equation (1), the first difference equation for $ln(proce)$ in Stigler's markup measure can be derived to

\[\Delta ln(p)= \alpha  + \Delta X' \times ln(\beta) + \epsilon. \tag{4}\]

The model, market variables and their corresponding coefficients have been combined into $X'$ in this equation. Rejecting the null hypothesis of $\alpha = 0$ indicates that change in price across products do not maintain equal markup, thus constituting price discrimination by our definition. 

\section{Results}

We collected data of 27 iPhone models in over fifty markets, and received 969 data points. The price of the iPhones are gathered from themacindex.com and the market information are mostly from World Bank. An overview of the model data is shown in Table 1. Summary and detailed data for market-specific variables are shown in Table 5 and 6 in Appendix. 


\begin{table}[H]
\centering
\begin{tabular}{|
>{\columncolor[HTML]{FFFFFF}}l |
>{\columncolor[HTML]{FFFFFF}}c |
>{\columncolor[HTML]{FFFFFF}}c |
>{\columncolor[HTML]{FFFFFF}}c |
>{\columncolor[HTML]{FFFFFF}}c |}
\hline
         & \textbf{Mean} & \textbf{Standard Deviation} & \textbf{Binary Variable} & \textbf{(Min, Max)} \\ \hline
iPhoneXR & 0.115         & 0.319                       & True                     & (0,   1)            \\ \hline
iPhone11 & 0.125         & 0.331                       & True                     & (0,   1)            \\ \hline
iPhone12 & 0.556         & 0.497                       & True                     & (0,   1)            \\ \hline
iPhone13 & 0.204         & 0.403                       & True                     & (0,   1)            \\ \hline
se       & 0.109         & 0.312                       & True                     & (0,   1)            \\ \hline
mini     & 0.105         & 0.307                       & True                     & (0,   1)            \\ \hline
pro      & 0.365         & 0.482                       & True                     & (0,   1)            \\ \hline
max      & 0.248         & 0.432                       & True                     & (0,   1)            \\ \hline
storage  & 273.24        & 204.60                      & False                    & (64, 1024)          \\ \hline
price    & 1214.35       & 442.22                      & False                    & (399,   2950)       \\ \hline
\end{tabular}
\caption{Summary Statistics of the dataset. Number of observations: 969.}
\end{table}

If Assumption 1 holds, we expect the listed price to be in constant ratio to the cost, which means we can examine the validity of Assumption 3 by testing how well we can estimate price of iPhones. We first regress the proposed cost function in equation (3), and the results are shown in Table 2.


\begin{table}[H]
\centering
\begin{tabular}{|l|c|c|}
\hline
                                        & \textbf{Adj R$^2$} & \textbf{Root MSE} \\ \hline
With Model-specific Controls            & 0.858          & 166.6             \\ \hline
With Model and Market-specific Controls & 0.924          & 121.8             \\ \hline
\end{tabular}
\caption{OLS regression statistics of the estimated cost using model and market-specific variables. Number of observations: 969.}
\end{table}

As expected, the Adj. $R^2$ and root MSE measures are both improved when market-specific variables are included with the model-specific ones. With 969 samples, the cost estimation function explains 92.4$\%$ percent of price variation in the original dataset, and the estimations are likely to be $\$$121.8 away from the true price. Considering the average price of iPhones in the dataset is $\$$1214.35, we consider this estimation accurate enough to keep Assumption 3. The Breusch Pagan test indicates heteroskedasticiy in the model; details of the regression along with robust standard errors are reported in Table 7 (Appendix). To identify what variables within the cost estimation model can be correlated with the error term, we regressed the squared-residuals with respect to the model and market-specific variables, and the coefficients on ``iPhone11”, ``iPhone12”, ``iPhone13”, ``storage”, ``tariff”, ``gdppc”, and the constant appear to be significant at 5$\%$. The squared-residual is smaller when ``iPhone11”, ``iPhone12”, or ``iPhone13” equal to 1; it increases as storage or tariff increases, and it decreases as gdppc increases. We suspect that “iPhone XR”, the base model, might be priced differently than the rest of the models, which can cause the error to increase when predicting price for an ``iPhone XR". The storage, tariff, and gdppc variables may also contain non-linear relationships to price; however, when we include storage$^2$, tariff$^2$, and gdppc$^2$ in the cost estimation function, the significance between these variables and the squared-residuals in the subsequent Bruesch Pagan test did not decrease. It is possible that these variables along with the presence of heteroskedasticity imply non-linearity in Apple’s pricing, which may indicate some sort of price discrimination. Next, we estimate the first difference in $ln(price)$ model derived from Stigler’s definition of price discrimination, the results are shown in Table 3.

\pagebreak
\begin{table}[h!]
\centering
\small
\begin{tabular}{|l|c|c|}
\hline
                                                                                                                       & \textbf{With Model-specific Controls}                                             & \begin{tabular}[c]{@{}c@{}}\textbf{With Model-specific and} \\ \textbf{Market-specific Controls}\end{tabular} \\ \hline
                                                                                                              $\alpha$         & \begin{tabular}[c]{@{}c@{}}0.460***\\ (0.014, 0.021)\end{tabular}        & \begin{tabular}[c]{@{}c@{}}0.188***\\ (0.015, 0.022)\end{tabular}                           \\ \hline
iPhone11                                                                                                               & \begin{tabular}[c]{@{}c@{}}0.068***\\ (0.019, 0.024)\end{tabular}        & \begin{tabular}[c]{@{}c@{}}0.070***\\ (0.014, 0.021)\end{tabular}                           \\ \hline
iPhone12                                                                                                               & \begin{tabular}[c]{@{}c@{}}0.322***\\ (0.018, 0.023)\end{tabular}        & \begin{tabular}[c]{@{}c@{}}0.323***\\ (0.013, 0.020)\end{tabular}                           \\ \hline
iPhone13                                                                                                               & \begin{tabular}[c]{@{}c@{}}0.278***  \\ (0.020, 0.024)\end{tabular}      & \begin{tabular}[c]{@{}c@{}}0.272*** \\ (0.015, 0.021)\end{tabular}                          \\ \hline
se                                                                                                                     & \begin{tabular}[c]{@{}c@{}}-0.607***\\ (0.018, 0.017)\end{tabular}       & \begin{tabular}[c]{@{}c@{}}-0.607***\\ (0.014, 0.011)\end{tabular}                          \\ \hline
mini                                                                                                                   & \begin{tabular}[c]{@{}c@{}}-0.120*** \\ (0.018, 0.016)\end{tabular}      & \begin{tabular}[c]{@{}c@{}}-0.121*** \\ (0.014, 0.010)\end{tabular}                         \\ \hline
pro                                                                                                                    & \begin{tabular}[c]{@{}c@{}}0.185***  \\ (0.018, 0.016)\end{tabular}      & \begin{tabular}[c]{@{}c@{}}0.185***  \\ (0.014, 0.009)\end{tabular}                         \\ \hline
max                                                                                                                    & \begin{tabular}[c]{@{}c@{}}0.075***\\ (0.018, 0.016)\end{tabular}        & \begin{tabular}[c]{@{}c@{}}0.075***\\ (0.013, 0.009)\end{tabular}                           \\ \hline
storage                                                                                                                & \begin{tabular}[c]{@{}c@{}}0.000623***\\ (2.73E-5, 2.91E-5)\end{tabular} & \begin{tabular}[c]{@{}c@{}}0.000624)***\\ (2.03E-5, 2.34E-5)\end{tabular}                   \\ \hline
\begin{tabular}[c]{@{}l@{}}tariff\\ (simple mean of manufaced\\  products tariff applied)\end{tabular}                 & \_                                                                       & \begin{tabular}[c]{@{}c@{}}0.018***   \\ (0.002, 0.002)\end{tabular}                        \\ \hline
\begin{tabular}[c]{@{}l@{}}salestax\\ (VAT for OECD countries)\end{tabular}                                            & \_                                                                       & \begin{tabular}[c]{@{}c@{}}0.014***\\ (5E-4, 4.9E-4)\end{tabular}                           \\ \hline
\begin{tabular}[c]{@{}l@{}}gdppc\\ (GDP per Capita in 2019/2020)\end{tabular}                                       & \_                                                                       & \begin{tabular}[c]{@{}c@{}}-5.29E-7* \\ (2.73E-7, 2.35E-7)\end{tabular}                     \\ \hline
\begin{tabular}[c]{@{}l@{}}consump\\ (household and NPISH final \\ Consumption expenditure per capita)\end{tabular} & \_                                                                       & \begin{tabular}[c]{@{}c@{}}1.52E-6**\\ (7.22E-7, 6.7E-7)\end{tabular}                       \\ \hline
Adj. R$^2$                                                                                                                 & 0.856                                                                    & 0.921                                                                                       \\ \hline
Breusch Pagan P-value                                                                                                  & 0.005                                                                    & 0.0000                                                                                      \\ \hline
\end{tabular}
\caption{Coefficients from regression price of iPhones. Coefficients are displayed with standard errors in parentheses in format (Ordinary SE, Robust SE); The asterisks *, **, and *** indicate significance levels at 10\%, 5\%, and 1\%, respectively. The left column contains regression results using just model-specific variables for estimation of price while the while column utilizes both model and market-specific controls. The model-specific controls, except for “storage”, are binary variables derived directly from the name of the phone. For example, if a model’s listed name is “iPhone 12 Pro 128gb”, its model-specific vector would contain (0,1,0,0,0,1,0,128). The data on tariff, gdppc, and consump are gathered from World Bank, and data on salestax are gathered from international Tax Competitiveness Index and Trading Economics. The base case used to calculate the differences is the iPhone XR with 64gb of storage.}
\end{table}

\pagebreak

All regressed variables in Table 3 are significant at 10 percent while most are significant at 1 percent. The effect of the variables on $\Delta ln(price)$ is as expected. The Pro and Max models are Apple’s premium models while the SE and Mini are the more affordable ones. The iPhone XR, 11, 12, and 13 are released in chronological order with each being more expensive than the previous one, except for the newly released iPhone 13, which is marginally cheaper than the previous year’s model. Among the market-specific variables, tariff and salestax are expected to increase listed price. The resultant coefficients estimate that one percentage point of increment in tariff or salestax will increase the listed price by $1.8 \%$ and \$1.4 \%, respectively. The null hypothesis that $\alpha=0$ implies that with differentiated cost considered, the various prices of iPhone models sold in different markets maintain constant profit markup. This null hypothesis is rejected at 0.1\% which implies that some sort of price discrimination may exist within iPhones’ pricing across different models and markets, we then attempt to determine whether socio-economical factors are significant to Apple’s differentiated prices.

\begin{table}[ht]
\centering
\small
\begin{tabular}{|l|c|c|c|c|}
\hline
\rowcolor[HTML]{FFFFFF} 
\cellcolor[HTML]{FFFFFF}                                                                              & \textbf{Model I.}                                                            & \textbf{Model II.}                                                               & \textbf{Model III.}                                                                & \textbf{Model IV.}                                                               \\ \hline
\begin{tabular}[c]{@{}l@{}}gini\\ (Gini Index)\end{tabular}                                           & \begin{tabular}[c]{@{}c@{}}12.863 ***\\ (2.897, 3.319)\end{tabular} &             -                                                            & \begin{tabular}[c]{@{}c@{}}9.128 ***\\ (3.056, 3.06)\end{tabular}         & \begin{tabular}[c]{@{}c@{}}9.568 ***\\ \\ (3.1, 3.146)\end{tabular}     \\ \hline
\begin{tabular}[c]{@{}l@{}}high10\\ (Percentage of   income held\\  by the highest 10\%)\end{tabular} & \begin{tabular}[c]{@{}c@{}}-7.07 *\\ (4.11, 4.21)\end{tabular}      &          -                                                               & \begin{tabular}[c]{@{}c@{}}-7.943 *\\ (4.409, 4.271)\end{tabular}         & \begin{tabular}[c]{@{}c@{}}-9.401 **\\ (4.62, 4.625)\end{tabular}       \\ \hline
\begin{tabular}[c]{@{}l@{}}ishighinc\\ (High Income Economy)\end{tabular}                             &                               -                                      & \begin{tabular}[c]{@{}c@{}}-43.657 **\\ (18.01, 19.444)\end{tabular}    & \begin{tabular}[c]{@{}c@{}}-35.737 *\\ (20.149,   22.854)\end{tabular}    & \begin{tabular}[c]{@{}c@{}}-29.878\\ (20.192,   22.236)\end{tabular}    \\ \hline
\begin{tabular}[c]{@{}l@{}}isuppermiddleinc\\ (Upper Middle Income\\ Economy)\end{tabular}              &                                     -                                & \begin{tabular}[c]{@{}c@{}}82.554 ***\\ (14.639,   19.475)\end{tabular} & \begin{tabular}[c]{@{}c@{}}72.352 ***\\ (15.363,   18.068)\end{tabular}   & \begin{tabular}[c]{@{}c@{}}71.262 ***\\ (15.328,   17.578)\end{tabular} \\ \hline
\begin{tabular}[c]{@{}l@{}}hhi\\ (Herfindahl-Hirschman \\Index)\end{tabular}                            &                                     -                                &            -                                                             & \begin{tabular}[c]{@{}c@{}}0.188 ***\\ (0.0517, 0.048)\end{tabular}       & \begin{tabular}[c]{@{}c@{}}0.165 **\\ (0.0667, 0.065)\end{tabular}      \\ \hline
hhi$^2$                                                                                                  &                                          -                           &          -                                                               & \begin{tabular}[c]{@{}c@{}}1.03E-5 ***\\ (3.56E-6,   3.4E-6)\end{tabular} & \begin{tabular}[c]{@{}c@{}}-4.73E-6\\ (4.3E-6,   4.13E-6)\end{tabular}  \\ \hline
\begin{tabular}[c]{@{}l@{}}iosshare\\ (Share of IOS)\end{tabular}                                     &                                -                                     &          -                                                               &                                          -                                 & \begin{tabular}[c]{@{}c@{}}11.835 ***\\ (4.456, 4.471)\end{tabular}     \\ \hline
iosshare$^2$                                                                                             &                                -                                     &         -                                                                &                            -                                               & \begin{tabular}[c]{@{}c@{}}-0.133 ***\\ (0.045, 0.044)\end{tabular}     \\ \hline
Breusch Pagan P-value                                                                                 & 0.0000                                                              & 0.0000                                                                  & 0.0000                                                                    & 0.0000                                                                  \\ \hline
RESET P-value                                                                                         & 0.0000                                                              & 0.0000                                                                  & 0.0000                                                                    & 0.0000                                                                  \\ \hline
\end{tabular}
\caption{Coefficients from four Regression Models. Coefficients are displayed with standard errors in parentheses in format (Ordinary SE, Robust SE); The asterisks *, **, and *** indicate significance levels determined with robust SE at 10\%, 5\%, and 1\%, respectively.}
\end{table}

 We include socio-economical factors of the market into the price prediction function, the model and market-specific variables are kept as controls. Table 4 displays the regression results. The significance of gini and high10 are significant across models I, III, and IV. The Gini Index here is on a scale of 0-100. The coefficients on gini indicate that for every 1\% point increase in gini, the price of an iPhone is likely to increase by around \$9. The variable high10 is not very significant across most models, but it indicates that iPhones are cheaper in countries where the top 10\% income group have more wealth. The markets are divided into high income economies, upper middle income economies, and middle income economies with the middle income markets being the base. The coefficients indicate that being an upper middle income economy likely increases iPhone prices by over \$70 dollars, which is consistently significant throughout the models. This relation is reversed for high income economies, but loses significance as more variables are added. 
 
 The Herfindahl-Hirschman Index (HHI) measures the competitiveness of a market; a lower HHI would indicate high market competitiveness while a high HHI indicates saturation. The hhi and hhi$^2$ are both significant in Model III, but their significance drop significantly when IOS market share variables are included. There is likely to be correlations between hhi and iosshare which means Model III. suffer from omitted variable bias. When both hhi and iosshare are present in Model IV, we see that as competition weakens, the price of iPhones increase. Keep in mind that the HHIs are in the magnitude of thousands, so the coefficient of 0.165 on hhi in Model IV. can potentially account for price difference of hundreds of dollars between competitive and saturated markets. We observe an significance second degree polynomial relationship between iosshare and price of iPhones in Model IV. It is estimated that for every percentage point of increase in share of IOS in a market, the price increases but at a diminishing rate.  The parabola has been graphed in Figure 1. The domain of iosshare in this dataset is between 2.91 (India) and 67.25 (Japan), which is within the domain where iosshare positively affects price. 
 
\vspace{10mm}
 
\begin{center}

\includegraphics[scale=0.3]{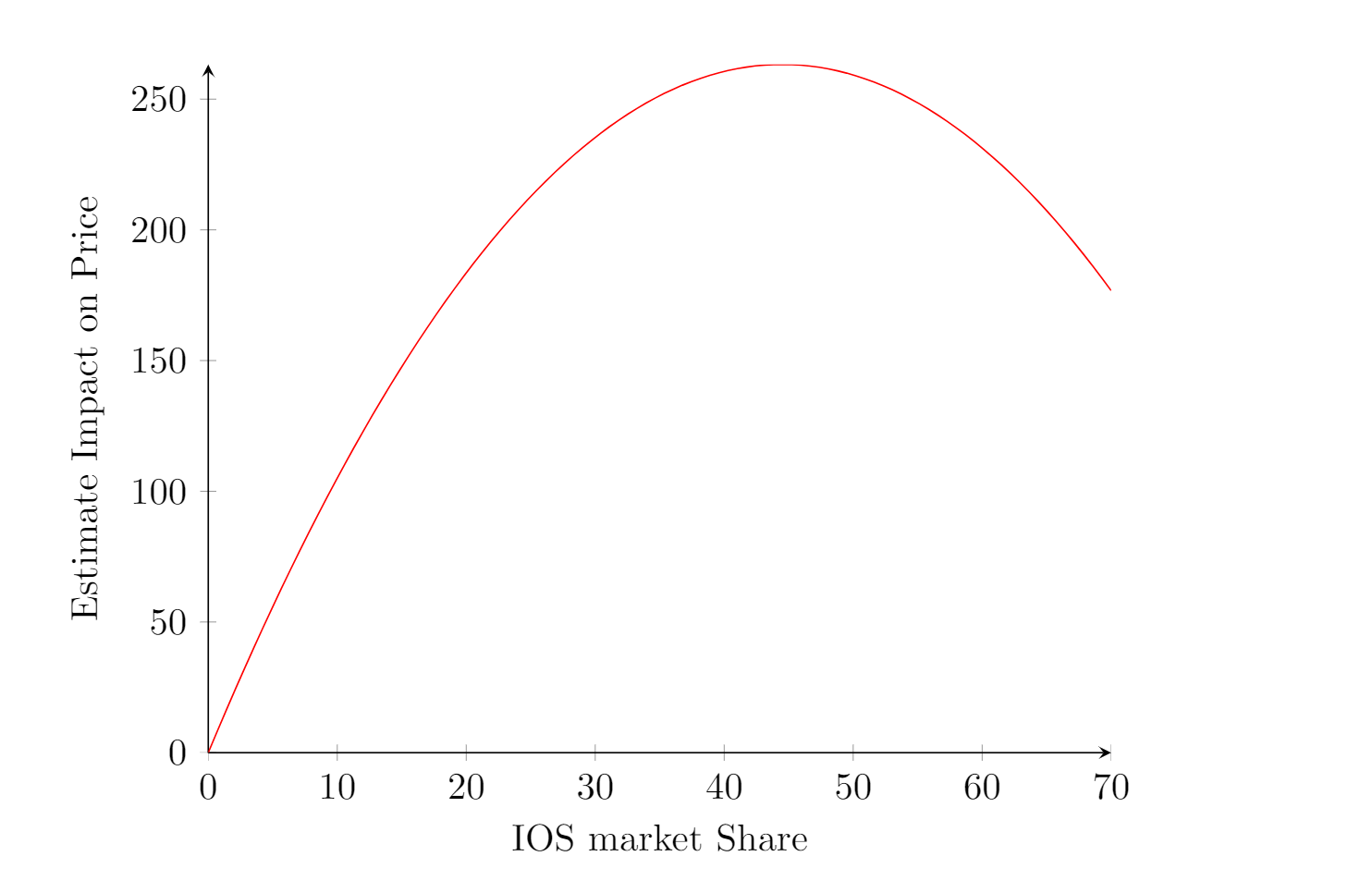}

\end{center}


Figure 1: The modeled relationship between IOS market share and price of iPhones across various markets.

\section{Discussion \& Conclusion}
In this paper, price discrimination in iPhones has been identified with Stigler’s definition on the basis of socio-economical factors. We find many variables to be both statistically and economically significant. For example, the market share of IOS can incur an increment of \$263.28 to an iPhone’s price when it is at 44.49\%. One unit of increment in Gini Index can increase price by over \$9. Our results also show that iPhones are cheaper in high income than middle income than upper middle income economies. This may imply that iPhones are generally more expensive in countries where wealth is less balanced. This is consistent with our hypothesis that the prices of iPhones are higher in markets where iPhones are deemed as premium or luxury goods. The classical theories on price discrimination agree that market power and lack of incentives for arbitrage lead to price discrimination. We are able to consider market power and competitiveness which will be discussed later in this section. By not including controls for arbitrage in our models, we do not deny the significance of arbitrage in identifying price discrimination. The cost of arbitrage is simply too difficult to define and quantify in this context. 

One persistent issue that occurs across this study is rejection of the null hypothesis when running the Breusch-Pagan test, which can indicate heteroskedasticity in the model. We believe heteroskedasticity can arise from both model-specific and market-specific variables. Our assumption that the cost of iPhones can be predicted linearly lead to the cost estimation model, which assumes that tags such “pro” has the same effect on an iPhone XR on an “iPhone 13”, but this may not be the case. This will cause heteroskedasticity; but keep in mind that this will not be correlated with any geographical variables, therefore will not affect conclusions of this study. Price discriminations can arise from price variation across products of different quality or across products sold at different geographical locations, and we are interested in the latter. Thus, heteroskedasticity that arise from market-specific variables are more important in this study. To discuss the presence of heteroskedasticity, we first propose that when more than two prices are being compared, identification of price discrimination using equation (4) will likely lead to heteroskedasticity. In equation (4), $\alpha$ symbolizes the difference in markups. When we are only comparing two prices, there is only one difference and $\alpha$ may be a constant. When we compare more than two prices, the value of $\alpha$ turns into a sample from a distribution and a constant can no longer represent such difference in markups any more. This means a portion of differences in markups, which is correlated with explanatory variables, is integrated with the error term, and we get heteroskedasticity. Table 9 (Appendix) indicates the error term is correlated with tariff and GDP per capita, these variables may very well serve as basis for price discrimination. Heteroskedasticity caused by model-specific variables are not necessarily relevant to this study, and heteroskedasticity caused by market-specific variables may be a result of price discrimination; So our solutions to heteroskedasticity is to maintain the original model and report the robust standard errors.

One may also argue that most of our prices are gathered from developed economies: 61\% of markets included in this study belong to high income economies. This is certainly a limitation to our study, but is also one that we could not avoid. In general, more iPhone models are available in high-income markets. The HHI and IOS market share are both found to be positively correlated with price of iPhones. This confirms the classical theory that market power is a requirement to price discrimination, and that lack of competition encourages price discrimination. It is worth noting that reverse causality may be present between price of iPhones and IOS market share. The price of iPhones directly influence sales and thus affects IOS market share. To address this issue requires additional instrumental variables, which may be done as an extension of this study. Including instrument variables for arbitrage may also be future work that stems from this study.

The purpose of this study is to identify and quantify socio-economical and geographical influences to iPhone prices. Our results confirms traditional theories on price discrimination and found economically significant factors. It is certainly unethical if a company provides higher prices to poorer markets to maximize profit, especially when the company’s mission statement is ``To make a contribution to the world by making tools for the mind that advance humankind”, according to the Economist. But price discrimination is not unlawful. After all, Apple is a public company pressured by the board of trustees to show revenue growth in its annual reports.

\bibliographystyle{plain} 
\bibliography{refs} 
\nocite{*}

\pagebreak

\section{Appendix}

\begin{table}[!htbp]
\centering
\begin{tabular}{|l|l|l|l|l|l|}
\hline
\textbf{Variable}       & \textbf{\# of Observations} & \textbf{Mean}     & \textbf{Std. Dev.} & \textbf{Min}     & \textbf{Max}      \\ \hline
gdppc          & 54              & 27785.75 & 25209.94  & 1155.1  & 115873.6 \\ \hline
gini           & 54              & 34.1     & 7.01      & 25      & 63       \\ \hline
tariff         & 54              & 3.15     & 2.87      & 0       & 14       \\ \hline
consump        & 54              & 14370.94 & 10852.34  & 606     & 40706    \\ \hline
high10         & 54              & 26.91    & 5.36      & 19.5    & 50.5     \\ \hline
povhead        & 54              & 1.7      & 4.36      & 0       & 22.5     \\ \hline
salestax       & 54              & 17.21    & 6.05      & 0       & 27       \\ \hline
highinc        & 54              & 0.61     & 0.49      & 0       & 1        \\ \hline
uppermiddleinc & 54              & 0.22     & 0.42      & 0       & 1        \\ \hline
middleinc      & 54              & 0.17     & 0.38      & 0       & 1        \\ \hline
iosshare       & 54              & 26.06    & 15.89     & 2.91    & 67.245   \\ \hline
ioshhi         & 54              & 6532.74  & 1139.57   & 4961.22 & 9080.31  \\ \hline
\end{tabular}
\caption{Summary statistics of the market-specific variables.}
\end{table}

\noindent \textbf{gdppc}: GDP per capita. Source: World Bank\\
\textbf{gini}: Gini Index. Source: World Bank\\
\textbf{tariff}: Simple mean of manufaced products tariff applied. Source: World Bank\\
\textbf{consump}: household and NPISH final Consumption expenditure per capita. Source: World Bank\\
\textbf{high10}: Income share held by highest 10\%. Source: World Bank\\
\textbf{povhead}: Poverty headcount ratio at \$1.9 a day. Source: World Bank\\
\textbf{salestax}: Sales Tax. Source: international tax competitiveness index \& trading economics.\\
\textbf{highinc}: Identified by World Bank as high income economy.\\
\textbf{uppermiddleinc}: Identified by World Bank as upper middle income economy.\\
\textbf{middleinc}: Identified by World Bank as middle income economy.\\
\textbf{Iosshare}: Share of IOS users in the market.  Source: StatCounter.\\
\textbf{ioshhi}: Herfindahl-Hirschman Index. Calculated using data from StatCounter.\\

\begin{table}[]
\centering
\tiny
\begin{tabular}{|
>{\columncolor[HTML]{FFFFFF}}l |
>{\columncolor[HTML]{FFFFFF}}l |
>{\columncolor[HTML]{FFFFFF}}l |
>{\columncolor[HTML]{FFFFFF}}l |
>{\columncolor[HTML]{FFFFFF}}l |
>{\columncolor[HTML]{FFFFFF}}l |
>{\columncolor[HTML]{FFFFFF}}l |
>{\columncolor[HTML]{FFFFFF}}l |
>{\columncolor[HTML]{FFFFFF}}l |
>{\columncolor[HTML]{FFFFFF}}l |
>{\columncolor[HTML]{FFFFFF}}l |
>{\columncolor[HTML]{FFFFFF}}l |
>{\columncolor[HTML]{FFFFFF}}l |}
\hline
country        & \multicolumn{1}{c|}{\cellcolor[HTML]{FFFFFF}gdppc} & \multicolumn{1}{c|}{\cellcolor[HTML]{FFFFFF}gini} & \multicolumn{1}{c|}{\cellcolor[HTML]{FFFFFF}tariff} & \multicolumn{1}{c|}{\cellcolor[HTML]{FFFFFF}consump} & \multicolumn{1}{c|}{\cellcolor[HTML]{FFFFFF}high10} & \multicolumn{1}{c|}{\cellcolor[HTML]{FFFFFF}povhead} & \multicolumn{1}{c|}{\cellcolor[HTML]{FFFFFF}\begin{tabular}[c]{@{}c@{}}sales\\tax\end{tabular}} & \multicolumn{1}{c|}{\cellcolor[HTML]{FFFFFF}\begin{tabular}[c]{@{}c@{}}high-\\inc \end{tabular}} & \multicolumn{1}{c|}{\cellcolor[HTML]{FFFFFF}\begin{tabular}[c]{@{}c@{}}upper-\\middle-\\inc \end{tabular}} & \multicolumn{1}{c|}{\cellcolor[HTML]{FFFFFF}\begin{tabular}[c]{@{}c@{}}midd-\\ leinc\end{tabular}} & \multicolumn{1}{c|}{\cellcolor[HTML]{FFFFFF}iosshare} & \multicolumn{1}{c|}{\cellcolor[HTML]{FFFFFF}ioshhi} \\ \hline
Australia      & 51812.2                                            & 34.4                                              & 2.1                                                 & 30552                                                & 27                                                  & 0.5                                                  & 10                                                                                                     & 1                                                    & 0                                                                                                       & 0                                                                                                  & 55.49                                                 & 4990.6927                                           \\ \hline
Austria        & 48105.4                                            & 30.8                                              & 1.8                                                 & 23019                                                & 23.9                                                & 0.6                                                  & 20                                                                                                     & 1                                                    & 0                                                                                                       & 0                                                                                                  & 30.625                                                & 5575.7416                                           \\ \hline
Belarus        & 6411.2                                             & 25.3                                              & 5.3                                                 & 4888                                                 & 21.3                                                & 0                                                    & 20                                                                                                     & 0                                                    & 1                                                                                                       & 0                                                                                                  & 13.81                                                 & 7437.5679                                           \\ \hline
Belgium        & 44594.4                                            & 27.2                                              & 1.8                                                 & 22209                                                & 22.2                                                & 0.1                                                  & 21                                                                                                     & 1                                                    & 0                                                                                                       & 0                                                                                                  & 36.2975                                               & 5361.9009                                           \\ \hline
Brazil         & 6796.8                                             & 53.3                                              & 14                                                  & 6743                                                 & 40.2                                                & 4.6                                                  & 17                                                                                                     & 0                                                    & 1                                                                                                       & 0                                                                                                  & 12.22                                                 & 7638.5577                                           \\ \hline
Canada         & 43241.6                                            & 33.3                                              & 1.5                                                 & 28220                                                & 25.3                                                & 0.2                                                  & 5                                                                                                      & 1                                                    & 0                                                                                                       & 0                                                                                                  & 52.2375                                               & 4965.988                                            \\ \hline
Chile          & 13231.7                                            & 44.4                                              & 1                                                   & 8644                                                 & 36.3                                                & 0.3                                                  & 19                                                                                                     & 1                                                    & 0                                                                                                       & 0                                                                                                  & 15.785                                                & 7498.8816                                           \\ \hline
China          & 10500.4                                            & 38.5                                              & 5.2                                                 & 3339                                                 & 29.4                                                & 0.5                                                  & 13                                                                                                     & 0                                                    & 1                                                                                                       & 0                                                                                                  & 21.7025                                               & 6835.0196                                           \\ \hline
Croatia        & 13828.5                                            & 29.7                                              & 1.8                                                 & 9011                                                 & 22.7                                                & 0.4                                                  & 25                                                                                                     & 1                                                    & 0                                                                                                       & 0                                                                                                  & 13.745                                                & 7953.7572                                           \\ \hline
Cyprus         & 26623.8                                            & 32.7                                              & 1.8                                                 & 19755                                                & 27.2                                                & 0                                                    & 19                                                                                                     & 1                                                    & 0                                                                                                       & 0                                                                                                  & 22.3125                                               & 6493.4033                                           \\ \hline
\begin{tabular}[c]{@{}l@{}}Czech \\ Republic  \end{tabular}            & 22762.2                                     & 25                                                & 1.8                                                 & 11051                                                & 21.5                                                & 0                                                    & 21                                                                                                     & 1                                                    & 0                                                                                                       & 0                                                                                                  & 19.925                                                & 6666.6377                                           \\ \hline
Denmark        & 60908.8                                            & 28.2                                              & 1.8                                                 & 29349                                                & 23.5                                                & 0.2                                                  & 25                                                                                                     & 1                                                    & 0                                                                                                       & 0                                                                                                  & 52.3575                                               & 5038.8947                                           \\ \hline
Egypt          & 3547.9                                             & 31.5                                              & 4.8                                                 & 2379                                                 & 26.9                                                & 3.8                                                  & 14                                                                                                     & 0                                                    & 0                                                                                                       & 1                                                                                                  & 11.475                                                & 7847.1135                                           \\ \hline
Estonia        & 23312.3                                            & 30.3                                              & 1.8                                                 & 10623                                                & 22.4                                                & 0.2                                                  & 20                                                                                                     & 1                                                    & 0                                                                                                       & 0                                                                                                  & 31.04                                                 & 5576.1933                                           \\ \hline
Finland        & 49041.3                                            & 27.3                                              & 1.8                                                 & 25058                                                & 22.6                                                & 0.1                                                  & 24                                                                                                     & 1                                                    & 0                                                                                                       & 0                                                                                                  & 27.91                                                 & 5908.748                                            \\ \hline
France         & 38615.1                                            & 32.4                                              & 1.8                                                 & 21950                                                & 26.7                                                & 0                                                    & 20                                                                                                     & 1                                                    & 0                                                                                                       & 0                                                                                                  & 31.7175                                               & 5824.9742                                           \\ \hline
Georgia        & 4278.9                                             & 35.9                                              & 0.8                                                 & 4140                                                 & 27.6                                                & 3.8                                                  & 18                                                                                                     & 0                                                    & 1                                                                                                       & 0                                                                                                  & 21.0425                                               & 6549.3361                                           \\ \hline
Germany        & 45723.6                                            & 31.9                                              & 1.8                                                 & 24206                                                & 24.6                                                & 0                                                    & 19                                                                                                     & 1                                                    & 0                                                                                                       & 0                                                                                                  & 29.86                                                 & 5679.252                                            \\ \hline
Greece         & 17676.2                                            & 32.9                                              & 1.8                                                 & 15170                                                & 24.9                                                & 0.1                                                  & 24                                                                                                     & 1                                                    & 0                                                                                                       & 0                                                                                                  & 13.1575                                               & 7386.4622                                           \\ \hline
Hungary        & 15899.1                                            & 29.6                                              & 1.8                                                 & 8869                                                 & 23.2                                                & 0.2                                                  & 27                                                                                                     & 1                                                    & 0                                                                                                       & 0                                                                                                  & 16.66                                                 & 6727.6703                                           \\ \hline
India          & 1900.7                                             & 35.7                                              & 8.9                                                 & 1101                                                 & 30.1                                                & 22.5                                                 & 18                                                                                                     & 0                                                    & 0                                                                                                       & 1                                                                                                  & 2.9075                                                & 9080.3126                                           \\ \hline
Indonesia      & 3869.6                                             & 38.2                                              & 6.2                                                 & 2377                                                 & 29.9                                                & 2.7                                                  & 10                                                                                                     & 0                                                    & 0                                                                                                       & 1                                                                                                  & 5.2                                                   & 8547.8537                                           \\ \hline
Ireland        & 83812.8                                            & 31.4                                              & 1.8                                                 & 22560                                                & 25.4                                                & 0.2                                                  & 23                                                                                                     & 1                                                    & 0                                                                                                       & 0                                                                                                  & 43.8875                                               & 5091.4367                                           \\ \hline
Israel         & 43610.5                                            & 39                                                & 1.4                                                 & 18472                                                & 27.8                                                & 0.2                                                  & 17                                                                                                     & 1                                                    & 0                                                                                                       & 0                                                                                                  & 22.6125                                               & 6468.5106                                           \\ \hline
Italy          & 31676.2                                            & 35.9                                              & 1.8                                                 & 19385                                                & 26.7                                                & 1.4                                                  & 22                                                                                                     & 1                                                    & 0                                                                                                       & 0                                                                                                  & 24.895                                                & 6185.3164                                           \\ \hline
Japan          & 40113.1                                            & 32.9                                              & 1.3                                                 & 27024                                                & 26.4                                                & 0.7                                                  & 10                                                                                                     & 1                                                    & 0                                                                                                       & 0                                                                                                  & 67.245                                                & 5352.4879                                           \\ \hline
\begin{tabular}[c]{@{}l@{}}Kazak- \\ hstan  \end{tabular}      & 9055.7                                             & 27.8                                              & 4.6                                                 & 5887                                                 & 23.5                                                & 0                                                    & 12                                                                                                     & 0                                                    & 0                                                                                                       & 1                                                                                                  & 21.1725                                               & 6676.858                                            \\ \hline
Latvia         & 17620                                              & 35.1                                              & 1.8                                                 & 9296                                                 & 26.9                                                & 0.3                                                  & 21                                                                                                     & 1                                                    & 0                                                                                                       & 0                                                                                                  & 25.5825                                               & 6179.8669                                           \\ \hline
\begin{tabular}[c]{@{}l@{}}Luxem- \\ bourg  \end{tabular}     & 115873.6                                           & 35.4                                              & 1.8                                                 & 31970                                                & 26.6                                                & 0.3                                                  & 17                                                                                                     & 1                                                    & 0                                                                                                       & 0                                                                                                  & 45.895                                                & 5093.8953                                           \\ \hline
Malaysia       & 10401.8                                            & 41.1                                              & 5.8                                                 & 6717                                                 & 31.3                                                & 0                                                    & 10                                                                                                     & 0                                                    & 0                                                                                                       & 1                                                                                                  & 17.39                                                 & 6582.9937                                           \\ \hline
Mexico         & 8346.7                                             & 45.4                                              & 3.1                                                 & 5985                                                 & 36.4                                                & 1.7                                                  & 16                                                                                                     & 0                                                    & 1                                                                                                       & 0                                                                                                  & 15.6525                                               & 7355.4853                                           \\ \hline
Moldova        & 4551.1                                             & 25.7                                              & 4.9                                                 & 3141                                                 & 22                                                  & 0                                                    & 20                                                                                                     & 0                                                    & 1                                                                                                       & 0                                                                                                  & 13.315                                                & 7484.3698                                           \\ \hline
Nepal          & 1155.1                                             & 32.8                                              & 12.6                                                & 606                                                  & 26.4                                                & 15                                                   & 15                                                                                                     & 0                                                    & 0                                                                                                       & 1                                                                                                  & 6.7425                                                & 8624.8983                                           \\ \hline
\begin{tabular}[c]{@{}l@{}}Nether- \\ lands  \end{tabular}     & 52304.1                                            & 28.1                                              & 1.8                                                 & 22163                                                & 23                                                  & 0.1                                                  & 21                                                                                                     & 1                                                    & 0                                                                                                       & 0                                                                                                  & 39.225                                                & 5183.256                                            \\ \hline
Norway         & 67294.5                                            & 27.6                                              & 0.3                                                 & 37546                                                & 22.2                                                & 0.3                                                  & 25                                                                                                     & 1                                                    & 0                                                                                                       & 0                                                                                                  & 52.6025                                               & 5054.6859                                           \\ \hline
Peru           & 6126.9                                             & 32.8                                              & 1.3                                                 & 3797                                                 & 31.1                                                & 2.2                                                  & 18                                                                                                     & 0                                                    & 1                                                                                                       & 0                                                                                                  & 8.1275                                                & 8325.3837                                           \\ \hline
Philippines    & 3298.8                                             & 42.3                                              & 3.4                                                 & 2098                                                 & 33.5                                                & 2.7                                                  & 12                                                                                                     & 0                                                    & 0                                                                                                       & 1                                                                                                  & 17.3975                                               & 7277.6911                                           \\ \hline
Poland         & 15656.2                                            & 30.2                                              & 1.8                                                 & 9806                                                 & 24                                                  & 0.2                                                  & 23                                                                                                     & 1                                                    & 0                                                                                                       & 0                                                                                                  & 3.2025                                                & 8891.6067                                           \\ \hline
Portugal       & 22439.9                                            & 33.5                                              & 1.8                                                 & 14893                                                & 26.6                                                & 0.3                                                  & 23                                                                                                     & 1                                                    & 0                                                                                                       & 0                                                                                                  & 23.33                                                 & 6464.4478                                           \\ \hline
Romania        & 12896.1                                            & 35.8                                              & 1.8                                                 & 8010                                                 & 24.9                                                & 2.4                                                  & 19                                                                                                     & 0                                                    & 1                                                                                                       & 0                                                                                                  & 17.645                                                & 6949.1804                                           \\ \hline
Russia         & 10126.7                                            & 37.5                                              & 5.3                                                 & 5936                                                 & 29.9                                                & 0                                                    & 20                                                                                                     & 0                                                    & 1                                                                                                       & 0                                                                                                  & 25.625                                                & 6254.728                                            \\ \hline
Slovakia       & 19156.9                                            & 25                                                & 1.8                                                 & 11069                                                & 19.5                                                & 0.2                                                  & 20                                                                                                     & 1                                                    & 0                                                                                                       & 0                                                                                                  & 18.7975                                               & 6704.4614                                           \\ \hline
\begin{tabular}[c]{@{}l@{}}South \\  Africa \end{tabular}     & 5090.7                                             & 63                                                & 7.5                                                 & 4264                                                 & 50.5                                                & 18.7                                                 & 15                                                                                                     & 0                                                    & 1                                                                                                       & 0                                                                                                  & 12.815                                                & 7336.7496                                           \\ \hline
\begin{tabular}[c]{@{}l@{}}South \\ Korea  \end{tabular}      & 31489.1                                            & 31.4                                              & 2.4                                                 & 13006                                                & 24                                                  & 0.2                                                  & 10                                                                                                     & 1                                                    & 0                                                                                                       & 0                                                                                                  & 25.33                                                 & 6332.7454                                           \\ \hline
Spain          & 27057.2                                            & 34.7                                              & 1.8                                                 & 15952                                                & 25.3                                                & 0.9                                                  & 21                                                                                                     & 1                                                    & 0                                                                                                       & 0                                                                                                  & 20.28                                                 & 6741.028                                            \\ \hline
Sri Lanka      & 3682                                               & 39.3                                              & 8.1                                                 & 2763                                                 & 32.6                                                & 0.9                                                  & 8                                                                                                      & 0                                                    & 0                                                                                                       & 1                                                                                                  & 8.4825                                                & 8062.1371                                           \\ \hline
Sweden         & 51925.7                                            & 30                                                & 1.8                                                 & 25893                                                & 22.9                                                & 0.7                                                  & 25                                                                                                     & 1                                                    & 0                                                                                                       & 0                                                                                                  & 53.5425                                               & 4977.8659                                           \\ \hline
\begin{tabular}[c]{@{}l@{}}Switzer- \\ land  \end{tabular}     & 86601.6                                            & 33.1                                              & 0.7                                                 & 40706                                                & 25.8                                                & 0                                                    & 7.7                                                                                                    & 1                                                    & 0                                                                                                       & 0                                                                                                  & 54.475                                                & 4961.2151                                           \\ \hline
Thailand       & 7189                                               & 34.9                                              & 7.3                                                 & 3314                                                 & 27.2                                                & 0.1                                                  & 7                                                                                                      & 0                                                    & 1                                                                                                       & 0                                                                                                  & 21.15                                                 & 6126.7236                                           \\ \hline
Turkey         & 8538.2                                             & 41.9                                              & 1.3                                                 & 9038                                                 & 31.6                                                & 0.4                                                  & 18                                                                                                     & 0                                                    & 1                                                                                                       & 0                                                                                                  & 16.905                                                & 7322.2835                                           \\ \hline
UAE            & 43103.3                                            & 26                                                & 4.1                                                 & 15847                                                & 20                                                  & 0                                                    & 5                                                                                                      & 1                                                    & 0                                                                                                       & 0                                                                                                  & 21.1475                                               & 6068.3613                                           \\ \hline
Ukraine        & 3726.9                                             & 26.6                                              & 1.9                                                 & 2821                                                 & 22.3                                                & 0                                                    & 20                                                                                                     & 0                                                    & 0                                                                                                       & 1                                                                                                  & 19.4125                                               & 6910.2928                                           \\ \hline
\begin{tabular}[c]{@{}l@{}}United \\ Kingdom  \end{tabular}   & 40284.6                                            & 35.1                                              & 1.8                                                 & 24819                                                & 26.7                                                & 0.3                                                  & 20                                                                                                     & 1                                                    & 0                                                                                                       & 0                                                                                                  & 49.745                                                & 4975.0528                                           \\ \hline
\begin{tabular}[c]{@{}l@{}}United \\ States  \end{tabular}    & 63543.6                                            & 41.4                                              & 0                                                   & 38594                                                & 30.8                                                & 1                                                    & 0                                                                                                      & 1                                                    & 0                                                                                                       & 0                                                                                                  & 55.87                                                 & 5166.6969                                           \\ \hline
\end{tabular}
\caption{Detailed market-specific parameters employed.}
\end{table}

\begin{table}[]
\centering
\begin{tabular}{|
>{\columncolor[HTML]{FFFFFF}}l 
>{\columncolor[HTML]{FFFFFF}}c 
>{\columncolor[HTML]{FFFFFF}}c 
>{\columncolor[HTML]{FFFFFF}}c 
>{\columncolor[HTML]{FFFFFF}}c 
>{\columncolor[HTML]{FFFFFF}}c 
>{\columncolor[HTML]{FFFFFF}}c |}
\hline
\multicolumn{7}{|c|}{\cellcolor[HTML]{FFFFFF}\textbf{Model Fit}}                                                                                                                                                                                                                                                                                                                                                                                                                                                          \\ \hline
\multicolumn{3}{|l|}{\cellcolor[HTML]{FFFFFF}\# of Observations}                                                                                                                                                                               & \multicolumn{4}{c|}{\cellcolor[HTML]{FFFFFF}969}                                                                                                                                                                                                                         \\ \hline
\multicolumn{3}{|l|}{\cellcolor[HTML]{FFFFFF}F(8,960)}                                                                                                                                                                                         & \multicolumn{4}{c|}{\cellcolor[HTML]{FFFFFF}732.5}                                                                                                                                                                                                                       \\
\hline
\end{tabular}
\caption{Regression Results of the estimated cos function.}
\end{table}

\begin{table}[]
\centering
\begin{tabular}{|
>{\columncolor[HTML]{FFFFFF}}l 
>{\columncolor[HTML]{FFFFFF}}c 
>{\columncolor[HTML]{FFFFFF}}c 
>{\columncolor[HTML]{FFFFFF}}c 
>{\columncolor[HTML]{FFFFFF}}c 
>{\columncolor[HTML]{FFFFFF}}c 
>{\columncolor[HTML]{FFFFFF}}c |}
\hline
\multicolumn{7}{|c|}{\cellcolor[HTML]{FFFFFF}\textbf{Model Fit}}                                                                                                                                                                                                                                                                                                                                                                                                                                                           \\ \hline
\multicolumn{3}{|l|}{\cellcolor[HTML]{FFFFFF}\# of Observations}                                                                                                                                                                               & \multicolumn{4}{c|}{\cellcolor[HTML]{FFFFFF}969}                                                                                                                                                                                                                          \\ \hline
\multicolumn{3}{|l|}{\cellcolor[HTML]{FFFFFF}F(8,960)}                                                                                                                                                                                         & \multicolumn{4}{c|}{\cellcolor[HTML]{FFFFFF}983.63}                                                                                                                                                                                                                       \\
\hline
\end{tabular}
\caption{Regression results with both model and market-specific variables.}
\end{table}

\begin{table}[]
\centering
\begin{tabular}{|
>{\columncolor[HTML]{FFFFFF}}l 
>{\columncolor[HTML]{FFFFFF}}c 
>{\columncolor[HTML]{FFFFFF}}c 
>{\columncolor[HTML]{FFFFFF}}c 
>{\columncolor[HTML]{FFFFFF}}c 
>{\columncolor[HTML]{FFFFFF}}c 
>{\columncolor[HTML]{FFFFFF}}c |}
\hline
\multicolumn{7}{|c|}{\cellcolor[HTML]{FFFFFF}\textbf{Model Fit}}                                                                                                                                                                                                                                                                                                                                                                                                                                                          \\ \hline
\multicolumn{3}{|l|}{\cellcolor[HTML]{FFFFFF}\# of Observations}                                                                                                                                                                               & \multicolumn{4}{c|}{\cellcolor[HTML]{FFFFFF}969}                                                                                                                                                                                                                         \\ \hline
\multicolumn{3}{|l|}{\cellcolor[HTML]{FFFFFF}F(8,960)}                                                                                                                                                                                         & \multicolumn{4}{c|}{\cellcolor[HTML]{FFFFFF}10.06}                                                                                                                                                                                                                       \\
\hline
\end{tabular}
\caption{Regression results of the squared-residuals in the cost estimation function with respect to model and market-specific variables}
\end{table}

\end{document}